\documentclass[aps,prd,twocolumn,showpacs,nofootinbib,amsmath,amssymb,amsfonts]{revtex4}
\usepackage{graphicx}

\newcommand{\dd}{\mathrm{d}}

\newcommand\beq{\begin{equation}}
\newcommand\eeq{\end{equation}}
\def\iso#1#2{\mbox{${}^{#2}{\rm #1}$}}

\def\he#1{\iso{He}{#1}}
\def\li#1{\iso{Li}{#1}}
\def\beryl#1{\iso{Be}{#1}}
\def\be#1{\iso{Be}{#1}}
\def\bore#1{\iso{B}{#1}}
\def\ga{\mathrel{\raise.3ex\hbox{$>$\kern-.75em\lower1ex\hbox{$\sim$}}}}
\def\la{\mathrel{\raise.3ex\hbox{$<$\kern-.75em\lower1ex\hbox{$\sim$}}}}

\begin{document}
%%%%%%%%%%%%%%%%%%%%%%%%%%%%%%%%%%%%%%%%%%%%%%%%%%%%%%%%%%%%%%%%%%%%%%%%%%%%%%%%%%%%%%%%

\leftline{UMN--TH--2526/06, FTPI--MINN--06/36}

\title{Coupled Variations of Fundamental Couplings and Primordial Nucleosynthesis}

\author{Alain Coc}
 \email{coc@csnsm.in2p3.fr}
 \affiliation{Centre de Spectrom\'etrie Nucl\'eaire et de
Spectrom\'etrie de Masse, IN2P3/CNRS/UPS,
B\^at. 104, 91405 0rsay Campus (France)}

\author{Nelson J. Nunes}
 \email[Electronic address: ]{nunes@damtp.cam.ac.uk}
 \affiliation{William I. Fine Theoretical Physics Institute,
              University of Minnesota, Minneapolis, MN 55455 (USA)}

\author{Keith A. Olive}
 \email{olive@physics.unm.edu}
 \affiliation{William I. Fine Theoretical Physics Institute,
              University of Minnesota, Minneapolis, MN 55455 (USA)}

\author{Jean-Philippe Uzan}
 \email{uzan@iap.fr}
 \affiliation{Institut d'Astrophysique de Paris,
              UMR-7095 du CNRS, Universit\'e Pierre et Marie
              Curie,
              98 bis bd Arago, 75014 Paris (France)}

\author{Elisabeth Vangioni}
 \email{vangioni@iap.fr}
 \affiliation{Institut d'Astrophysique de Paris,
              UMR-7095 du CNRS, Universit\'e Pierre et Marie
              Curie,
              98 bis bd Arago, 75014 Paris (France)}
\date{\today}
%%%%%%%%%%%%%%%%%%%%%%%%%%%%%%%%%%%%%%%%%%%%%%%%%%%%%%%%%%%%%%%%%%%%%%%%%%%%%%%%%%%%%%%%%
\begin{abstract}
The effect of variations of the fundamental nuclear parameters on big-bang 
nucleosynthesis are modeled and discussed in detail taking into account the interrelations between the
fundamental parameters arising in unified theories.
Considering only \he4, strong
constraints on the variation of the neutron lifetime, neutron-proton mass difference are set. These
constraints are then translated into constraints on the time variation of the Yukawa couplings and the fine structure constant. Furthermore, we show that
a variation of the deuterium binding energy is able to reconcile the \li7
abundance deduced from the WMAP analysis with its spectroscopically determined value
while maintaining concordance with D and \he4.
The analysis is strongly based on the potential model of Flambaum and Shuryak that relates the binding energy of the deuteron with the nucleon and $\sigma$ and $\omega$ mesons masses, however, we show that an alternative approach that consists of a pion mass dependence necessarily leads to equivalent conclusions.
\end{abstract}

\pacs{PACS}
\maketitle
%%%%%%%%%%%%%%%%%%%%%%%%%%%%%%%%%%%%%%%%%%%%%%%%%%%%%%%%%%%%%%%%%%%%%%%%%%%%%%%%%%%%%%%%%

\section{Introduction}

Primordial nucleosynthesis (BBN) is one of the most successful predictions of the standard hot
big-bang model. Its success rests on the concordance between the observational determinations of
the light element abundances of D and \he4 specifically, and their theoretically predicted
abundances \cite{bbn, bbnb}. Furthermore, measurements of the CMB anisotropies by WMAP \cite{wmap}
have led to precision determinations of the baryon density or equivalently the baryon-to-photon
ratio, $\eta$. As $\eta$ is the sole parameter of the standard model of BBN, it is possible to
make very accurate predictions \cite{cfo3,Coc04,cuoco,cyburt} and hence infer the expected
theoretical abundances of all of the light elements (including \li7, \li6, \beryl9, \beryl{10} and \bore{11}).

At present, a discrepancy between the predicted abundance of \li7 and its spectroscopically
determined abundance persists.
The \li7 abundance based on the WMAP baryon density is predicted to
be~\cite{Coc04}:
\begin{equation}
{\rm \li7/H} = 4.15^{+0.49}_{-0.45} \times 10^{-10}. \label{li7c}
\end{equation}
The systems best suited for Li observations are metal-poor halo stars in our Galaxy.  Analyses of
the abundances in these stars yields \cite{rbofn} ${\rm Li/H} = (1.23^{+0.34}_{-0.16}) \times
10^{-10}$ and more recently \cite{bona} ${\rm Li/H} = (1.26 \pm 0.26) \times
10^{-10}$. This value is in clear contradiction with most estimates of the primordial Li
abundance, as also shown by~\cite{cyburt} who find
\begin{equation}
{\rm \li7/H} = 4.26^{+0.73}_{-0.60} \times 10^{-10} \label{li7}.
\end{equation}
In both cases,  the \li7 abundance is a factor of $\sim 3$ higher than the value observed in most
halo stars.

There have been several attempts to account for the discrepancy between the
BBN/WMAP predicted value of \li7/H and its observational determination.
These include depletion mechanisms due to rotationally-induced mixing and/or diffusion.
Current estimates for possible depletion factors are in the range
$\sim$~0.2--0.4~dex~\cite{dep}.
However, the negligible intrinsic spread in Li \cite{rnb} leads to the conclusion
that depletion in halo stars is as low as 0.1~dex. It is also possible that the stellar
parameters used to determine the Li abundance from the spectroscopic measurements
may be systematically off.  Most important among these is the effective temperature assumed
for stellar atmospheres.
These can differ by up
to 150--200~K, with higher temperatures resulting in estimated Li abundances which are higher by
$\sim 0.08$~dex per 100~K.  Thus accounting for a difference of 0.5 dex between BBN and the
observations, would require a serious offset of the stellar parameters. We note that there has
been a recent analysis \cite{mr} which does support higher temperatures, and brings the
discrepancy between theory and observations to 2$\sigma$.

Another potential source for systematic uncertainty lies in the BBN calculation of the \li7
abundance. The predictions for \li7 carry the largest uncertainty of the 4 light elements which
stem from uncertainties in the nuclear rates. The effect of changing the yields of certain BBN
reactions was recently considered in Ref.~\cite{Coc04}.  In particular, they concentrated on the set of
cross sections which affect \li7 and are poorly determined both experimentally and theoretically.
It was found for example, that an increase of the $^7$Be$(d,p)2\he4$ reaction by a factor of 100
would reduce the \li7 abundance by a factor of about 3 in WMAP $\eta$ range. This reaction has
since been remeasured and precludes this solution \cite{Ang05}. The  possibility of systematic
errors in the $\he3(\alpha,\gamma) {}^7$Be reaction, which is the only important \li7 production
channel in BBN, was considered in detail in~\cite{cfo4}. However, the agreement between the
standard solar model and solar neutrino data  provides  constraints on variations in this cross
section.  Using the standard solar model of Bahcall~\cite{bah}, and recent solar neutrino
data~\cite{sno}, one can exclude systematic variations of the magnitude needed to resolve the BBN
\li7 problem at the $\ga 95\%$ CL~\cite{cfo4}. The ``nuclear fix'' to the \li7 BBN problem is
unlikely.

On the other hand, various theoretical explanations involving physics beyond the standard model
have been proposed~\cite{cyburt2}. One possible extension of the standard BBN scenario allows for
inhomogeneous nucleosynthesis~\cite{inhomobbn} but this seems to overproduce \li7. It has also
been argued that particle decay after BBN could lower the \li7 abundance and produce some \li6 as
well \cite{jed}. This has been investigated in the framework of the constrained minimal
supersymmetric standard model if the lightest supersymmetric particle is assumed to be the
gravitino \cite{susy}. Some models have been found which accomplish these goals \cite{Jed2}.
Another route is to assume that gravity is not described by general relativity but is attracted
toward general relativity during the cosmic evolution~\cite{dn}. BBN has been extensively studied
in that scenario (see e.g. Ref.~\cite{dp}). The effect of the modification of gravity is mainly to
induce a time variation of the equivalent speed-up that can be tuned to happen during BBN but it
can have other signatures both on cosmological and local scales~\cite{stother}.

In this article, we want to investigate the possible variations of
fundamental constants. It is well known that variations in the fundamental coupling constants such
as the fine structure constant, $\alpha$, can affect the light element abundance during BBN.
Most analyses have concentrated on the effect of such variations on the abundance of \he4
\cite{bbna,bbnb2,cyburt2,co,flam1,fl1,vvar,scherrer,lvar,otherbd}. Changes in $\alpha$ directly induce
changes in the nucleon mass, $\Delta m_N$, which affects the neutron-to-proton ratio.

Much of the recent excitement over the possibility of a time variation in the fine structure
constant stems from a series of recent observations of quasar absorption systems and a detailed
fit of positions of the absorption lines for several heavy elements using the ``many-multiplet''
method~\cite{webb,murphy3}. When this method is applied to a set of Keck/Hires data, a
statistically significant trend for a variation in $\alpha$ was reported: $\Delta \alpha / \alpha
= (-0.54 \pm 0.12) \times 10^{-5}$ over a redshift range $0.5 \la z \la 3.0$.  The minus sign
indicates a smaller value of $\alpha$ in the past. In Ref.~\cite{chand}, a set of  high signal-to-noise
systems yielded the result $\Delta \alpha / \alpha = (-0.06 \pm 0.06) \times 10^{-5}$. One should
note that both results are sensitive to assumptions regarding the isotopic abundances of elements
used in the analysis which further complicates the interpretation of any positive signal from
these analyses \cite{amo}.

In addition to the possible variation in $\alpha$,
it is reasonable to search for other time-varying quantities such as the
ratio of the proton-to-electron mass, $\mu$ \cite{mu0}.
Indeed recent analyses \cite{mu} claim  to observe a variation
at the level
\beq
\frac{\Delta \mu}{\mu} = (2.4 \pm 0.6) \times 10^{-5}
\label{muobs}
\eeq
using Lyman bands of H$_2$ spectra in two quasars.
In the following, we investigate the effect on BBN of changes in fundamental couplings which
could also account for a variation in $\mu$.  As we will see, the largest effect can be traced to
a variation of the Higgs  vacuum expectation value leading to a variation in
the binding energy of deuterium.  We find that for a suitably large variation  (of order  a few
hundredths of a percent) in $\mu$ at the time of BBN, the \li7 abundance can be decreased
by the requisite factor without overly affecting the agreement between theory and
observations for D and \he4.

Although we do not directly tie our calculations to the observation with the result in
Eq.~(\ref{muobs}), we do take as our starting point, the possibility that $\mu \equiv m_p/m_e$
could have differed from its present value at the time of BBN.  
As a result we will be interested in
variations of Yukawa couplings, $h$ (we can assume, or not, that all Yukawas vary identically);
the Higgs vacuum expectation value (vev), $v$, and $\Lambda \equiv \Lambda_{QCD}$.  
Some effects of
the variations of $v$ \cite{vvar,scherrer} and $\Lambda$ \cite{lvar} on BBN have been considered in the
past. Here we will be primarily interested in the interdependence of these variations (some of
which are {\em not} model dependent) and their effects on quantities of direct importance to BBN,
such as the binding energy of the deuteron, the nucleon mass difference and the neutron lifetime.

In complete generality, the effect of varying constants on BBN predictions is difficult to model
because of the intricate structure of QCD and its role in low energy nuclear reactions (see
Refs.~\cite{ctes,landau}). The abundances of light nuclei produced during BBN mainly depend on the
value of a series of fundamental constants which include, the gravitational constant $G$, the
three gauge couplings and the Yukawa couplings of the electron and quarks.
One needs to relate the nuclear parameters (cross-sections, binding energies and the masses of the
light nuclei) to these fundamental constants. This explains why most studies are restricted to a
subset of constants, such as e.g. the gravitational constant~\cite{dp,cyburt2,bbng}, the fine
structure constant~\cite{cyburt2,bbna,bbnb2,co} or $v$ \cite{vvar,scherrer}.

The approach we adopt here recognizes that many variations of parameters we deem fundamental are
interrelated in model dependent ways \cite{co}.  For example, if one assumes gauge coupling
unification at some high energy (grand unified) scale, there is a direct relation between
variations in the fine structure constant and $\Lambda$. In string theories, the variations of
gauge couplings will be related to variations in Yukawa couplings, and in models where the weak
scale is determined by dimensional transmutation \cite{weaktran}, there is a relation between
variations in the Yukawa couplings and variations in the Higgs vev.  Variations in the latter will
also trigger variations in $\Lambda$. While the exact relation between these variations is model
dependent, the fact that they are interrelated is not.  Therefore it is inconsistent for example
to consider a variation in $v$ without simultaneously varying $\Lambda$. We will make use of these
dependencies to study variations in several (tractable) quantities which affect BBN. Coupled
variations of this type were used to strengthen existing bounds on the
fine structure constant based on Oklo and meteoritic data \cite{opqccv}.  As noted
above, we can not fully evolve the variations in all nuclear reactions, because their dependence
is unknown.  Here, we will be primarily interested in the induced variations of the nucleon mass
difference, the neutron life-time, and the binding energy of deuterium. We recognize that this
represents a limitation on our results.

In Sec.~\ref{sec2}, we relate the variation of the BBN parameters, mainly $Q$, $\tau_n$ and
$B_D$, to the variation of the fundamental parameters such as the Yukawa couplings, $h$, the QCD
energy scale, $\Lambda$, and the fine structure constant, $\alpha$. Section~\ref{sec3} focuses on
the relations that can be drawn between the variations of the fundamental parameters, taking into
account successively grand unification, dimensional transmutation and the possibility that the
variation is driven by a dilaton. In order to deal with some of the theoretical uncertainties, we
introduce two phenomenological parameters and we then make the connection with the variation of
the proton to electron mass ratio at low redshift. Section~\ref{sec4} focuses on the BBN
computation and first describes the implementation of the variations in our BBN code. Assuming
that the fine structure constant does not vary we show that deuterium and \he4 data set strong
constraints on the variation of the Yukawa couplings [see Eq.~(\ref{limit})] but that inside this
bound there exists a range reconciling the \li7 abundance with spectroscopic observations. We
then allow the fine structure constant to vary and set a sharp constraint on its variation in the
dilaton scenario [see Eqs.~(\ref{limit2a}) and (\ref{limit2b})].

\section{From fundamental parameters to BBN quantities}\label{sec2}

As discussed above, we focus our attention on three physical quantities which have direct bearing
on the resulting abundances from BBN, the nucleon mass difference $Q = m_n - m_p = 1.29$~MeV, the
neutron lifetime $\tau_n$, and the binding energy of deuterium $B_D$.

The neutron-proton mass difference is expressed in terms of $\alpha, \Lambda, v$, and the $u$ and
$d$ quark Yukawa couplings as \beq Q \equiv m_n-m_p =  a \, \alpha \, \Lambda + (h_d-h_u) \,v \,,
\eeq where the electromagnetic contribution at present is $a \, \alpha_0 \, \Lambda_0 = -0.76$,
and therefore the weak contribution is $({h_d}_0-{h_u}_0)\, v_0 = 2.05$ \cite{pn}. The variation
of $Q$ will then scale as
\begin{equation}
\frac{\Delta Q}{Q} = -0.6\left[ \frac{\Delta \alpha}{\alpha}+ \frac{\Delta \Lambda}{\Lambda}
\right] + 1.6 \left[\frac{\Delta (h_d-h_u)}{h_d-h_u} + \frac{\Delta v}{v} \right] \,. \label{Q1}
\end{equation}
The neutron lifetime can be well approximated by
\begin{eqnarray}\label{eq:tau}
 \tau_n^{-1} &=& \frac{1}{60} \, \frac{1+3\, g_A^2}{2\pi^3} \, G_F^2 \, m_e^5 \,
                 \left[\sqrt{q^2-1} (2q^4-9q^2-8) \right. \nonumber \\
             &~& \left. + 15 \ln(q+\sqrt{q^2-1}) \right] \,,
\end{eqnarray}
where $q = Q/m_e$. Since $G_F =  1/\sqrt{2} v^2$ and $m_e = h_e v$ we have for the relative
variation of the neutron lifetime,
\begin{eqnarray}
\label{eq:dtau}
\frac{\Delta \tau_n}{\tau_n} &=& -4.8 \, \frac{\Delta v}{v} + 1.5 \, \frac{\Delta h_e}{h_e} - 10.4 \, \frac{\Delta (h_d-h_u)}{h_d-h_u} \nonumber \\
&~& + 3.8 \, \left(\frac{\Delta \alpha}{\alpha} + \frac{\Delta \Lambda}{\Lambda} \right) \,.
\label{tau1}
\end{eqnarray}

In addition to $Q$ and $\tau_n$, which have been well studied in the context of BBN, we consider
the variation of $B_D$. This is one of the better known quantities in the nuclear domain: it is
experimentally measured to a precision better than $10^{-6}$~\cite{Audi} so that allowing a change
of its value by a few \% at BBN can only be reconciled with laboratory measurements if its value
is varying with time.

Recently, in a series of works~\cite{flam1,fl1,fl} Flambaum and collaborators have considered the
dependence of hadronic properties on quark masses and have set constraints on the deuterium
binding energy from BBN~\cite{fl1} following Refs.~\cite{vvar,scherrer,lvar,otherbd}. The importance of
$B_D$ can be understood  by the fact that the equilibrium abundance of deuterium and the reaction
rate $p(n,\gamma)$D depend exponentially on $B_D$ and on the fact that the deuterium is in a
shallow bound state.

Here, we follow Refs.~\cite{fl1,fl} to compute the quark-mass dependence of the deuterium binding
energy. Using a potential model, the dependence of $B_D$ on the nucleon mass and $\sigma$ and
$\omega$ meson masses have been determined \cite{fl}
\begin{eqnarray}
\label{Bdflambaum}
 \frac{\Delta B_D}{B_D} = -48 \, \frac{\Delta m_\sigma}{m_\sigma} +
    50 \, \frac{\Delta m_\omega}{m_\omega}
    + 6 \, \frac{\Delta m_N}{m_N} \,, \label{flbd}
\end{eqnarray}
for constant $\Lambda$. One can see that the coefficients of the quantities on the RHS of (\ref{flbd}) do not add up to unity as it is required on dimensional grounds. Clearly there is a variation of a dimensional quantity that has not been taken into account, which at low energy, can be expressed in terms of an a priori unknown combination of $\Lambda$ and $v$. For definiteness we will write the missing term in terms of $\Lambda$ only, keeping in mind that this somewhat artificial method of fixing units is accounted for in the uncertainty in the relations between the variations of the fundamental parameters that we will discuss in the next section.
Hence, when the nucleon and meson masses are kept constant, we write $\Delta B_D/B_D = -7 \Delta \Lambda/\Lambda$.
On the other hand, fixing $\Lambda$, when varying quark masses (the
largest contribution comes from $m_s$), their result is $\Delta B_D/B_D = -17 \, \Delta m_s/m_s$.

The importance of the strange quark in the nucleon and meson masses can be traced to the
$\pi$-nucleon $\Sigma$ term, which is given by
\begin{equation}
  \sigma_{\pi N} \equiv \Sigma = {1 \over 2}
  (m_u + m_d) (B_u + B_d) .
\end{equation}
where
\begin{equation}\label{defbq}
 B_q \equiv  \langle p |  \bar{q} q | p \rangle \,
\end{equation}
Defining $y = 2B_s/(B_u + B_d)$, the combination $\Sigma (1-y)$ is the change in the nucleon mass
due to the non-zero $u, d$ quark masses, which is estimated on the basis of octet baryon mass
differences to be $\sigma_0 = 36 \pm 7$ MeV~\cite{oldsnp}.  Following Ref.~\cite{Cheng}, we have
$(B_u - B_s)/(B_d - B_s) = 1.49$ and given a value of $\Sigma$, one can determine $B_q$. In
Ref.~\cite{fl}, the value $B_s = 1.5$ was adopted and corresponds to $\Sigma \simeq 51$ MeV, which
is a reasonable value. This corresponds to
$$
\frac{\Delta m_N}{m_N} = \left(\frac{m_s B_s}{m_N}\right)\frac{\Delta m_s}{m_s}
 \simeq 0.19\frac{\Delta m_s}{m_s}.
$$
For these values we find a similar value (though slightly larger than the one found in Ref.~\cite{fl}) for the light quark ($u$ and $d$) contributions which give
$$
\frac{\Delta m_N}{m_N} \simeq 0.052 \frac{\Delta m_q}{m_q}.
$$
This implies that
\begin{equation}\label{varmp}
 \frac{\Delta m_p}{m_p}\simeq 0.76 \frac{\Delta\Lambda}{\Lambda}+0.24\left(\frac{\Delta h}{h}
 +\frac{\Delta v}{v}\right) \,.
\end{equation}

The value of $\Sigma$ however has substantial uncertainties which were recently discussed in
Ref.~\cite{eoss8}. A often used value is $\Sigma = 45$ MeV which was already somewhat larger than
naive quark model estimates, and corresponded to $y \simeq 0.2$. However, recent determinations of
the $\pi$-nucleon $\Sigma$ term have found higher values \cite{highsnp}, $\Sigma = 64$ MeV. Still
higher values can be ascertained for the observation of exotic baryons \cite{exotic}. For $\Sigma$
= 45 (64) MeV, $B_s$ = 0.9 (2.8) and $\Delta m_N/m_N$ = 0.12 (0.36) $\Delta m_s/ m_s$.  The
contribution from $u$ and $d$ quarks is 0.046 and 0.066 for $\Sigma =$ 45, 64 MeV, respectively. A similar calculation
for the $\omega$ meson leads to $\Delta m_\omega / m_\omega$ = (0.09, 0.15, 0.29) $\Delta m_s / m_s$ for
$\Sigma$ = 45, 51, and 64 MeV respectively.  For the $\sigma$ meson, three contributions were
identified in \cite{fl}, only one of which is related to $\Sigma$, yielding $\Delta m_\sigma /
m_\sigma$ = (0.44, 0.54, 0.75) $\Delta m_s / m_s$. Combining these sensitivities using Eq.~(\ref{flbd}), we would arrive at $\Delta B_D / B_D$ $= (-16, -17, -19)$ $\Delta m_s / m_s$ (when the
$u$ and $d$ contributions are neglected).  Thus, despite the large uncertainties in the individual
sensitivities, the dependence of $B_D$ on the strange quark mass is relatively stable. Because of
the cancellations in Eq.~(\ref{flbd}), the $u$ and $d$ quark contributions are indeed small:
$\Delta B_D / B_D$ $= (0.08, -0.009, -0.20)$ $\Delta m_q / m_q$ and can safely be neglected.

Choosing the central value $\Sigma = 51$ MeV and since $m_s  = h_s v$, we immediately have the relation between $B_D$, $h$, and $v$. Adding these
two contributions and using $\Delta B_D / B_D = -17 \Delta m_s / m_s$ we have in general,
\begin{equation}
\frac{\Delta B_D}{B_D} = 18\, \frac{\Delta \Lambda}{\Lambda} - 17 \,
\left(\frac{\Delta v}{v} + \frac{\Delta h_s}{h_s} \right) \,,
\label{bd1}
\end{equation}
where once again we have repaired the mass dimension by adding the appropriate powers
of $\Delta \Lambda/ \Lambda$.  

Eqs.~(\ref{Q1}), (\ref{tau1}), and (\ref{bd1}) form the initial basis for our computation.

\section{Relations between fundamental parameters}\label{sec3}

\subsection{General relations in a GUT}

We note that several relations among our fundamental parameters can be found.  First, changes in
either $h$ or $v$ trigger changes in $\Lambda$ \cite{svz}.  This is evident from the low energy
expression for $\Lambda$ when mass thresholds are included
\begin{equation}
\Lambda = \mu \left(\frac{m_c \, m_b \, m_t}{\mu^3} \right)^{2/27} \, \exp\left(-\frac{2\pi}{9\alpha_s(\mu)} \right) \,.
\end{equation}
for $\mu > m_t$ up to some unification scale in the standard model\footnote{In supersymmetric
models, additional thresholds related to squark and gluino masses would affect this relation \cite{Dent}.}
\begin{equation}
\label{DeltaLambda}
\frac{\Delta \Lambda}{\Lambda} = R \, \frac{\Delta \alpha}{\alpha} +
\frac{2}{27} \left(3 \, \frac{\Delta v}{v} + \frac{\Delta h_c}{h_c} + \frac{\Delta h_b}{h_b}
+  \frac{\Delta h_t}{h_t} \right) \,.
\end{equation}
The value of $R$ is determined by the particular grand unified theory and particle content which
control both the value of $\alpha(M_{GUT}) = \alpha_s(M_{GUT})$ and the low energy relation
between $\alpha$ and $\alpha_s$, leading to significant model dependence  in $R$ \cite{Dent,dine}.
Here we will assume a value of $R = 36$ corresponding to a set of minimal assumptions
\cite{co,Langacker}. However, in most BBN computations, we
will neglect the variation in $\alpha$ and therefore the precise value of
$R$ chosen will not affect our conclusions. Nevertheless, the relation between $h,v$ and $\Lambda$
is quite robust and has been neglected in most studies  discussing the effect of varying $v$ (or
varying $G_F$) \cite{vvar,scherrer}.

For the quantities we are interested in, we now have
\begin{eqnarray}
\frac{\Delta B_D}{B_D} &=& -13 \left(\frac{\Delta v}{v} + \frac{\Delta h}{h}\right)
                           + 18R \, \frac{\Delta \alpha}{\alpha}  \,, \\
\frac{\Delta Q}{Q} &=& 1.5 \left(\frac{\Delta v}{v} + \frac{\Delta h}{h}\right)
                       -0.6(1+R) \, \frac{\Delta \alpha}{\alpha} \,, \\
\frac{\Delta \tau_n}{\tau_n} &=& -4 \, \frac{\Delta v}{v} - 8\, \frac{\Delta h}{h}
                        + 3.8(1+R)\,\frac{\Delta \alpha}{\alpha} \,.
\end{eqnarray}
where we have assumed that all Yukawa couplings vary identically, $\Delta h_i/h_i = \Delta h/h$.
For clarity, we have written only rounded values of the coefficients, however, the numerical
computation of the light element abundances uses the more precise values. We also recall that $\Delta
G_F/G_F=-2\Delta v/v$ and $\Delta m_e/m_e=\Delta h/h+\Delta v/v$.

\subsection{Interrelations between fundamental parameters}

Secondly, in all models in which the weak scale is determined by dimensional transmutation,
changes in the largest Yukawa coupling, $h_t$, will trigger changes in $v$ \cite{weaktran}. In
such cases, the Higgs vev is derived from some unified mass scale (or the Planck scale)  and can
be written as (see Ref.~\cite{co})
\begin{equation}
v = M_P \exp\left( -\frac{8 \pi^2 c }{h_t^2}\right) \,,
\label{vmp}
\end{equation}
where $c$ is a constant of order unity. Indeed, in supersymmetric models with unification
conditions such as the constrained minimal supersymmetric standard model \cite{cmssm}, there is in
general a significant amount of sensitivity to the Yukawa couplings and the top quark Yukawa in
particular.  This sensitivity can be quantified by a fine-tuning measure defined by \cite{EENZ}
 \beq 
  \Delta_i \equiv {\partial \ln m_W\over \partial \ln a_i} 
 \eeq
where $m_W$ is the mass of the
$W$ boson and can be substituted with $v$. The $a_i$ are the input parameters of the supersymmetric
model and include $h_t$. In regions of the parameter space which provide a suitable dark matter
candidate \cite{eos}, the total sensitivity $\Delta = \sqrt{\sum_i \Delta_i^ 2}$ typically ranges
from 100 -- 400 for which the top quark contribution is in the range $\Delta_t = 80 - 250$. In
models where the neutralino is more massive, $\Delta$ may surpass 1000 and $\Delta_t$ may be as
large as $\sim 500$.

Clearly there is a considerable model dependence in the relation between $\Delta v$ and $\Delta
h_t$.  Here we assume a relatively central value obtained from Eq.~(\ref{vmp}) with  $c \simeq h_0
\simeq 1$. In this case we have
\begin{equation}
 \frac{\Delta v}{v} = 16  \pi^2 c \, \frac{\Delta h}{h^3}
                   \simeq 160 \, \frac{\Delta h}{h} \,,
\end{equation}
but in light of the model dependence,  we will set
\begin{equation}
\label{Deltav} \frac{\Delta v}{v} \equiv S\,  \frac{\Delta h}{h} \,,
\end{equation}
hence defining $S\equiv \dd\ln v/\dd\ln h \sim \Delta_t$ and keeping in mind that $S\simeq160$. It follows that
the variations of $B_D$, $Q$ and $\tau_n$ are expressed in the following way
\begin{eqnarray}
\frac{\Delta B_D}{B_D} &=& -17(S+1) \frac{\Delta h}{h}
                           + 18\, \frac{\Delta \Lambda}{\Lambda} \,, \\
\frac{\Delta Q}{Q} &=& 1.6(S+1) \frac{\Delta h}{h}
                   -0.6 \, \left( \frac{\Delta \alpha}{\alpha} + \frac{\Delta \Lambda}{\Lambda} \right) \,, \\
\frac{\Delta \tau_n}{\tau_n} &=&\!\!\! -(8.8+4.8S) \frac{\Delta h}{h}
                  \!+\! 3.8 \left(\frac{\Delta \alpha}{\alpha} + \frac{\Delta \Lambda}{\Lambda} \right)
\end{eqnarray}
where we have again assumed common variations in all of the Yukawa couplings. It also follows that
$\Delta G_F/G_F=-2S\Delta h/h$ and $\Delta m_e/m_e=(1+S)\Delta h/h$.

Now, using the relation~(\ref{DeltaLambda}) we arrive at
\begin{eqnarray}
\label{DeltaBd3}
 \frac{\Delta B_D}{B_D} &=& -13 (1+S) \, \frac{\Delta h}{h}
            + 18R \, \frac{\Delta \alpha}{\alpha} \, \\
\label{DeltaQ3}
 \frac{\Delta Q}{Q} &=& 1.5(1+S) \, \frac{\Delta h}{h}
            - 0.6(1+R) \, \frac{\Delta \alpha}{\alpha} \,, \\
\label{Deltatau3}
 \frac{\Delta \tau_n}{\tau_n} &=& -(8+4S)\frac{\Delta h}{h}
           + 3.8(1+R)\frac{\Delta \alpha}{\alpha} \,.
\end{eqnarray}

Finally we can take into account the possibility that the variation of the constants is induced by
an evolving dilaton~\cite{co}. In this scenario, it was shown that $\Delta h/h = (1/2) \Delta
\alpha/\alpha$, therefore the expressions above can be simplified to
\begin{eqnarray}
\label{DeltaBd4}
 \frac{\Delta B_D}{B_D} &=& -[6.5(1+S)-18R] \frac{\Delta \alpha}{\alpha} \, \\
\label{DeltaQ4}
 \frac{\Delta Q}{Q} &=& (0.1+0.7S-0.6R) \frac{\Delta \alpha}{\alpha} \, \\
\label{Deltatau4}
 \frac{\Delta \tau_n}{\tau_n} &=& -[0.2+2S-3.8R] \frac{\Delta \alpha}{\alpha} \,,
\end{eqnarray}
though these relations will also be affected by model dependent
threshold corrections.

\subsection{Sensitivity of $B_D$ to the pion mass}
An independent calculation suggests a large dependence of the binding energy of the deuteron to the pion mass \cite{epelbaum} parametrized in Ref.~\cite{scherrer}, for constant $\Lambda$, by
\begin{equation}
\label{Bdscherrer}
\frac{\Delta B_D}{B_D} = -r \, \frac{\Delta m_\pi}{m_\pi} \,,
\end{equation}
where $r$ is a fitting parameter found to be between 6 and 10. The mass of the pion is given by $f_{\pi}^2 m_{\pi}^2 = (m_u+m_d)\langle \bar{q} q \rangle$, where $f_\pi \propto \Lambda$ is a coupling and $\langle \bar{q} q \rangle \propto \Lambda^3$ is the quark condensate. Hence, the sensitivity of the binding energy to the fundamental parameters is
\begin{equation}
\label{bd1a}
\frac{\Delta B_D}{B_D} = \left(1+\frac{r}{2}\right)\frac{\Delta\Lambda}{\Lambda}  -\frac{r}{2}\left(\frac{\Delta v}{v}+\frac{\Delta h}{h}\right) \,,
\end{equation}
which must be compared with Eq.~(\ref{bd1}). The coefficients are different, at most, by a factor of 4. Substituting relations (\ref{DeltaLambda}) and (\ref{Deltav}) into Eq.~(\ref{bd1a}) we obtain
\begin{equation}
\frac{\Delta B_D}{B_D} =  (0.2-0.4 r)(1+S) \, \frac{\Delta h}{h} + (1+0.5r) 
R \, \frac{\Delta \alpha}{\alpha} \,,
\end{equation}
which is to be compared with Eq.~(\ref{DeltaBd3}). 

For the dilaton model considered, $\Delta h/h = (1/2) \Delta \alpha /\alpha$, and we find
\begin{equation}
\frac{\Delta B_D}{B_D} =   [(0.1-0.2r) (1+S) + (1+0.5r)R] \, \frac{\Delta \alpha}{\alpha} \,  \,.
\end{equation}
Again, the ratio of the coefficients in Eq.~(\ref{DeltaBd4}) to these is at the most of order 4.
We therefore conclude that taking a scaled dependence with the pion mass (\ref{Bdscherrer}) leads to the same constraints on the variation of the fundamental parameters (up to a factor of a few) as the nucleon/meson mass dependence (\ref{Bdflambaum}).

\subsection{Links to the variation of $m_p/m_e$}

Before we use the above relations in our BBN code, it is interesting to first compare these
relations with the observed variation in $\mu$. Using Eq.~(\ref{varmp}), and then
Eqs.~(\ref{DeltaLambda}) and (\ref{Deltav}), the proton-to-electron mass ratio,
$\mu = m_p/m_e$
varies according to
\begin{equation}
\frac{\Delta \mu}{\mu}  = 0.8 R\,\frac{\Delta\alpha}{\alpha}-0.6(S+1) \, \frac{\Delta h}{h} \,.
\end{equation}
Using the current value on the observational variation of $\mu$ at redshift $z \sim 3$~\cite{mu},
i.e. $\Delta\mu/\mu \approx 3 \times 10^{-5}$ we obtain, assuming $\alpha$ constant,
\begin{equation}
\frac{\Delta h}{h} \simeq -3.2 \times 10^{-7} \left(\frac{161}{1+S}\right) \,.
\end{equation}
Interestingly we deduce from Eq.~(\ref{DeltaBd3}) that when $\alpha$ is constant
\begin{equation}
\label{estimateBd1}
\frac{\Delta B_D}{B_D} \simeq 22 \frac{\Delta\mu}{\mu}\simeq 6.6 \times 10^{-4} 
\,,
\end{equation}
at $z \sim 3$, independent of the value of $S$. 

In the case where the variation is driven by a dilaton, we can link the observational variation in $\mu$ to a variation in $\alpha$ to get
\begin{equation}
 \frac{\Delta\alpha}{\alpha} = -1.5 \times10^{-6}\, \left[\frac{-20.2}{0.8 R-0.3
 (S+1)}\right]\,,
\end{equation}
which is compatible with the measurement of the time variation of the fine structure constant in Refs.~\cite{webb,murphy3} but
higher than the stronger bound found in Ref.~\cite{chand}, for the considered value $(R,S)=(36,160)$. Note that this
corresponds to $\Delta B_D/B_D\simeq 6 \times10^{-4}$, by applying
Eq.~(\ref{DeltaBd4}), which is comparable to the value found in Eq.~(\ref{estimateBd1}) where $\alpha$ was taken to be constant.

\section{Nucleosynthesis}\label{sec4}

\subsection{Implementation in a BBN code}

We incorporate the relations derived above in a BBN network. We use the reaction rates provided by
the compilation and analysis of experimental data of Ref.~\cite{Des04} covering ten of the twelve
nuclear reactions involved in the BBN. The two remaining reactions of importance,
$n\leftrightarrow p$ and $p(n,\gamma)$D come from theory and are numerically evaluated, taking
into account the variations discussed above.

The $p(n,\gamma)$D reaction rate is calculated according to the Chen and Savage~\cite{Che99}
derivation of the cross section in the framework of effective field theory. The weak reaction
rates $n\leftrightarrow p\,$ are calculated by numerical integration of the electron, positron and
neutrino Fermi distributions and phase space for the six weak interaction reactions. Zero
temperature radiative corrections~\cite{Dic82,Lop99} to the weak rates are also evaluated
numerically. The neutrino versus photon temperature used in these rate calculations is a byproduct
of the numerical integration of the Friedmann equation where we follow the electron-positron
annihilation exactly by the numerical integration of their Fermi distributions.

In principle, one could have
introduced a variation in $B_D$ in the effective field theory cross section provided the scattering
length in the $^1S_0$ channel follows this variation. We found it simpler to use the Dmitriev {\it
et al.}~\cite{fl1} prescription for the reaction rate change that takes into account these effects.
The binding energy of the deuteron is also directly involved in the calculation of the reverse
rate.

Variations in the binding energy of the dineutron also have little
effect on the primordial  abundances provided its absolute value remains smaller than the deuteron's binding energy \cite{kneller2004}.
Considering that, in this work the variations on the binding energy of
the deuteron are only of a few percent, we do not expect any important role played by the binding energy of the dineutron in the calculations.

By varying $B_D$ one also changes the
size of the deuteron which consequently modifies the D$({\rm D},n)^3 {\rm H_e}$,  D$({\rm D},p)$T and D$(p,\gamma)^3 {\rm H_e}$ reaction rates \cite{Rupak}. We have, however, not included this effect in our calculation.

The $n\leftrightarrow p$ weak rates depend on both $Q$ and $\tau_n$. Eq.~(\ref{eq:tau}) gives the
decay rate of free neutrons; that is, the zero temperature limit of the weak $n\rightarrow p$
rate. This equation is used, in conjunction with the experimental value of the neutron lifetime
($\tau_n=885.7$~s; see Ref.~\cite{PDG06}) to fix the normalization of the finite temperature weak
rates. Hence, variations in $\tau_n$ directly affect those rates\footnote{While a more
recent determination of neutron lifetime has been published \cite{newn}, its value 
has not been adopted by the RPP \cite{PDG06}.  Consequences of this 
measurement on BBN were considered in \cite{mks}.}. More precisely, we use
Eq.~(\ref{eq:tau}) to fix the {\em present day} normalization, then scale it according to the
values of $m_e$ and $G_F$ at BBN and use the BBN values of  $Q$ and $m_e$ to evaluate the
integrals in the finite temperature weak rate calculations involving $Q/m_e$, $T/m_e$ and
$T_\nu/m_e$. Nevertheless, the approximation of scaling the weak rates with $\tau_n$ is very
good.

\subsection{Results under the hypothesis $\Delta\alpha/\alpha=0$}

We proceed with the calculation of the light element abundances during nucleosythesis using
Eqs.~(\ref{DeltaBd3}), (\ref{DeltaQ3}) and (\ref{Deltatau3}) {\em first} neglecting the
contribution from $\Delta \alpha/\alpha$. 
From the discussions of the two previous sections, we conclude that variations in the Yukawa couplings lead directly to the changes in $\tau_n$, $Q$, and $B_D$. In addition, the nucleon mass also
changes ($\Delta m_N/ m_N \simeq 0.76 \Delta \Lambda/ \Lambda + 0.24 (\Delta v/ v + \Delta h/h) \simeq
66 \Delta h / h$) as does the electron mass ($\Delta m_e / m_e \simeq 161 \Delta h / h$).

\begin{figure}[htb]
\center
\includegraphics[width=9cm]{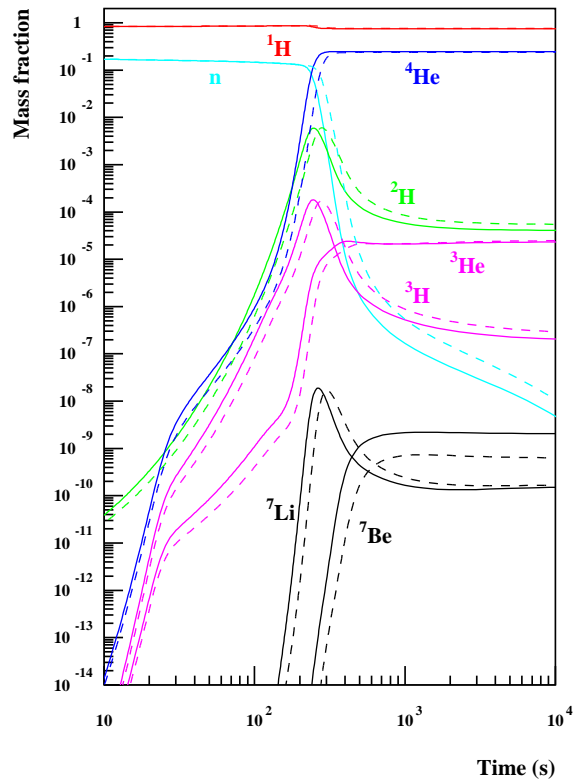}
\caption{The time evolution of the light element abundances for standard BBN (solid curves)
 and when variations $\Delta h/ h = 1.5\times 10^{-5}$ are included (dashed), assuming $\alpha$
 remains constant.
 The WMAP value of $\eta = 6.12 \times 10^{-10}$ was assumed and we have used $S=160$.}
\label{f:xfact}
\end{figure}

In Fig.~\ref{f:xfact}, we show the time evolution of the light elements for the standard BBN model
(solid curves) and when a variation  $\Delta h/ h = 1.5\times 10^{-5}$ (dashed curves) is considered,
assuming the WMAP value for $\eta$ and using the central value $S=160$. We see that the largest
effect is indeed a decrease in the \be7 abundance (which contributes to the final \li7 abundance
after decay) correlated to an increase of $n/p$ and a slight increase in the D/H abundance.

\begin{figure}[htb]
\center
\includegraphics[width=8cm]{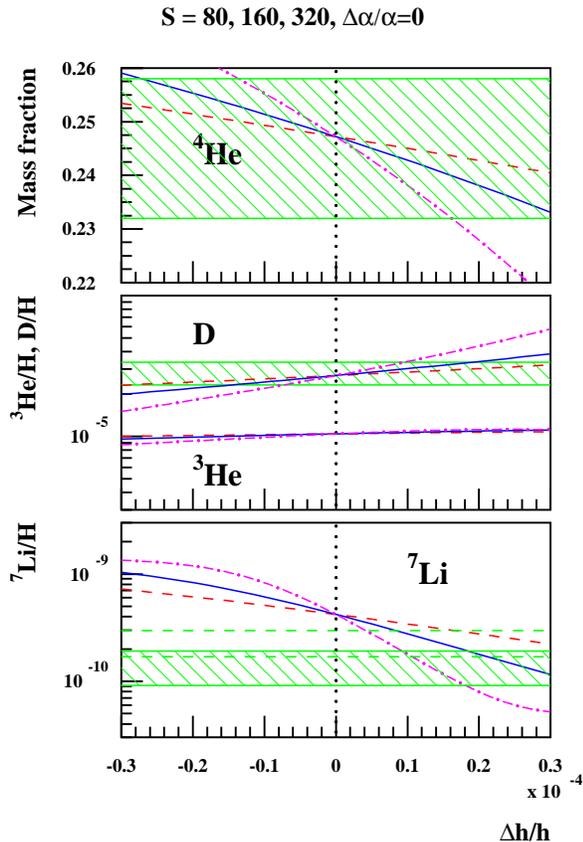}
\caption{Primordial abundances of \he4, D, \he3 and \li7  when allowing a variation of the
Yukawa couplings. The horizontal cross-hatched regions depict the $2\sigma$ spectroscopic data.
We have assumed 3 values for $S$: $S=80$ (dashed lines), $S=160$ (solid lines) and $S=320$ (dot-dashed lines) for $\eta = 6.12 \times 10^{-10}$.}
 \label{f:30pc}
\end{figure}

In Fig.~\ref{f:30pc}, we show the final abundances of D/H, \he3, \he4, and \li7/H as a function
of the variation in the Yukawa coupling $h$, for three assumed values of the parameter $S$.  The horizontal cross-hatched regions indicate the current observational
spectroscopic determinations.  For \he4, we use $Y = 0.232 - 0.258$ \cite{os}. For D/H, we use 2$\sigma$ range
based the latest average of six quasar absorption systems, D/H = $(2.83 \pm 0.52)\times10^{-5}$ \cite{omear}. For
\li7/H, we show two ranges: the first given by \li7/H = $1.23^{+0.68}_{-0.32} \times 10^{-10}$
\cite{rbofn} and the second given by \li7/H = $(2.33 \pm 0.64) \times 10^{-10}$ when higher surface
temperatures are assumed \cite{mr} and is represented with dashed lines. The dotted vertical line indicates the standard BBN results
(i.e. $\Delta h / h = 0$) for $\eta = 6.12 \times 10^{-10}$.

We recall that there is significant model dependence in several of the assumed relations between
the fundamental parameters. For example, in Eq.~(\ref{Deltav}), we adopted $c/h^2 = 1$ (that is
$S\simeq160$). However, the origin of the dependence between $v$ and $h$ depends on physics beyond
the standard model, and $c/h^2$ could be significantly larger or smaller than unity. 

As one can see from  Fig.~\ref{f:30pc}, each of the light elements, D, \he4, and \li7 show strong
dependences on $\Delta h / h$.  In fact, D/H provides us with the strongest constraint (under the
hypothesis that $\alpha$ is constant) which for $S = 160$ is,
\begin{equation}\label{limit}
 - 1.5 \times 10^{-5} <  \frac{\Delta h}{h} < 1.9 \times 10^{-5}.
\end{equation}
Using Eq.~(\ref{DeltaBd3}), this bound translates to $-4\times10^{-2} < \Delta B_D/B_D < 3.1 \times10^{-2}$.
Note that we have not used the \li7 abundance to set the lower bound on $\Delta h / h$. However,
we also observe that for values of $\Delta h / h \ga 1.8 \times 10^{-5}$
($\Delta h/h \ga 0.9 \times 10^{-5}$ for the second range of observational \li7), the \li7 abundance is
sufficiently small so as to come into agreement with the observational data. So long as we do not
exceed the upper bound given in Eq.~(\ref{limit}), all of the light elements can be brought into
agreement with data. Thus we must saturate the limit, but recall that this conclusion is based
under the restrictive assumption that $\alpha$ is constant. On the other hand, it also means that our hypothesis can be falsified by decreasing the error bars of either \li7 or D.

\begin{figure}[htb]
\center
\includegraphics[width=8cm]{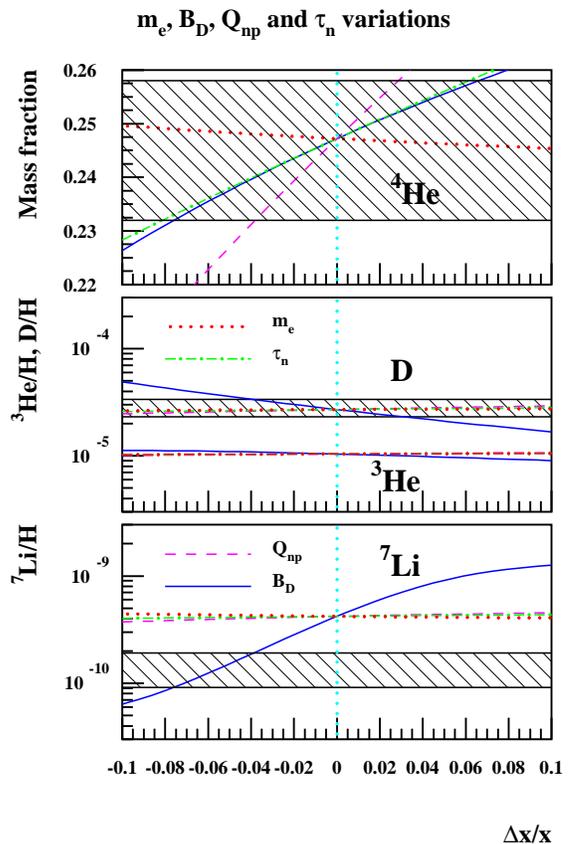}
\caption{Primordial abundances of the light nuclei as a function of the relative variation of
$m_e$ (dotted lines), $\tau_n$ (dot-dashed lines), $Q$ (dashed lines), and $B_D$ (solid lines)
with the same conventions as in Fig.~\ref{f:30pc}.} \label{f:var_all}
\end{figure}

In Fig.~\ref{f:var_all}, we show the individual contributions of the varying BBN quantities to the light element abundances.
When varying $Q$ and $m_e$ individually, $\tau_n$ is held constant, i.e. Eqs.~(\ref{eq:tau}) and (\ref{eq:dtau}) have not been used.
The effects of the variations of $\tau_n$, $Q$, $B_D$, and $m_e$ can be seen explicitly.
The curves for $m_e$ are due to the effects of the electron mass on the
expansion of the universe: $m_e$ effectively enters in the r.h.s. of the Friedmann equation,
affecting the timing and magnitude of the photon bath reheating following electron-positron
annihilations. This effect is however very small as seen in Fig.~\ref{f:var_all}. The effect of varying $m_e$ in the weak rates is accounted for in the overall variation of $\tau_n$.
The electron mass does not affect the abundance of any of the isotopes, however,
$\tau_n$ and $Q$ have a significant effect on \he4 leaving deuterium and
\li7 almost unchanged.

From the \he4 data, we deduce the bounds,
$-7.5\times10^{-2} \lesssim \Delta B_D/B_D \lesssim 6.5 \times 10^{-2}$, $-8.2\times10^{-2} \lesssim \Delta \tau_n/\tau_n \lesssim 6\times10^{-2}$ and $-4\times10^{-2} \lesssim \Delta Q/Q \lesssim 2.7 \times10^{-2}$
at $2\sigma$. A variation of the deuterium binding energy affects all the abundances, in particular, the deuterium data sets the tighter constraint
$-4\times10^{-2} \lesssim \Delta B_D/B_D \lesssim 3\times10^{-2}$. Interestingly, these bounds are equivalent to the ones obtained from the constraint (\ref{limit}) considering the interrelations between the fundamental parameters.
The \li7 abundance is brought in
concordance with spectroscopic observations provided its change falls within the interval
$-7.5\times10^{-2} \lesssim \Delta B_D/B_D \lesssim -4\times10^{-2}$.
We thus conclude that $B_D$ is the most important parameter connected to the discrepancy of the
\li7 abundance, and again, we see that there exists a window allowing for consistent \li7 and deuterium abundances with data.

One may also consider the effect of the variation of the nucleon mass. The proton and neutron
reduced mass enters as a factor $(m_p^{-1}+m_n^{-1})^{1\over2}$ in the $p(n,\gamma)$D rate. For
variations of the order we are considering, this effect is negligible.

\subsection{Allowing for $\Delta\alpha/\alpha\not=0$}

\begin{figure}[htb]
\center
\includegraphics[width=8cm]{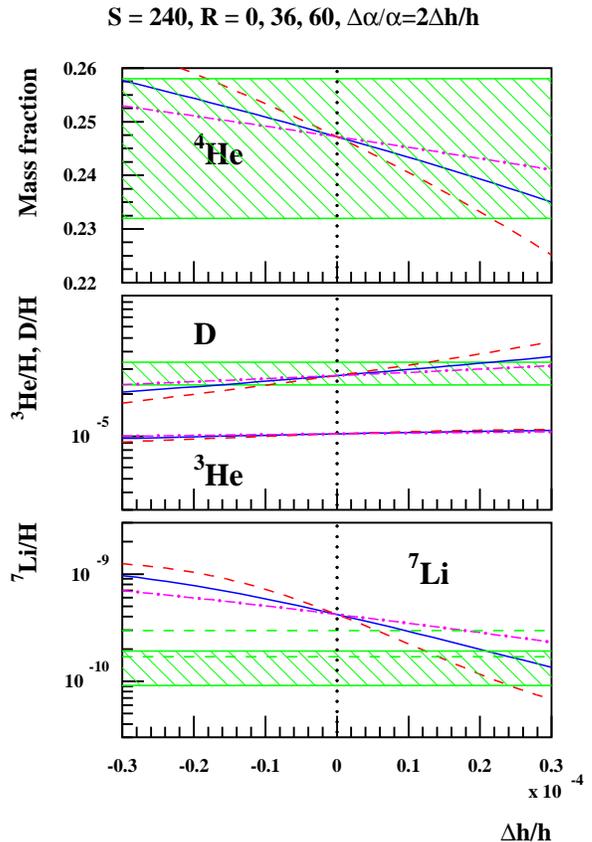}
\caption{Primordial abundances of \he4, D, \he3 and \li7 as a function of
$\Delta h/h=(1/2)\Delta\alpha/\alpha$  when allowing a variation of the
fine structure constant for three values of the $R$ parameter: $R=0$ (dashed lines), $R=36$
(solid lines) and $R=60$ (dot-dashed lines).}
 \label{alpha}
\end{figure}

We now allow the fine structure constant to vary and we further assume that it is tied to the variation of the
Yukawa couplings according to $\Delta h/h = (1/2) \Delta \alpha/\alpha$, using
Eqs.~(\ref{DeltaBd4})--(\ref{Deltatau4}). The results are shown in Fig.~\ref{alpha} where the abundances are depicted for three values of the parameter $R$. Comparison of this figure with Fig.~\ref{f:30pc} shows the
effect of including the variation in $\alpha$. Not considering \li7, the tighter bounds on $\Delta h/h$ are again given by the deuterium abundance and are comparable in 
order of magnitude to the ones found in Eq.~(\ref{limit}):
\begin{equation}
\label{limit2a}
-1.6 \times 10^{-5} <  \frac{\Delta h}{h} < 2.1 \times 10^{-5} \,,
\end{equation}
for $R = 36$ and
\begin{equation}
\label{limit2b}
-3 \times 10^{-5} <  \frac{\Delta h}{h} < 4 \times 10^{-5} \,,
\end{equation}
for $S = 240$ and $R = 60$.

While these limits are far more stringent than the one found in Ref.~\cite{bbna}, it is consistent with those derived in Refs.~\cite{bbnb2,co} where
coupled variations were considered. Once again, for a variation near
the upper end of the
range (\ref{limit2a}) and (\ref{limit2b}), we can simultaneously fit all of the observed abundances.

As noted above, a variation of $\alpha$ induces a multitude of changes in nuclear cross sections
that have not been included here. We have checked, however, that a variation of $\Delta
\alpha/\alpha \approx 4 \times 10^{-5}$ leads to variations in the reaction rates (numerically
fit), mainly through the Coulomb barrier, of the most important $\alpha$-dependent reactions in
BBN~\cite{bbna} that never exceed one tenth of a percent in magnitude.

\begin{figure}[htb]
\center
\includegraphics[width=8cm]{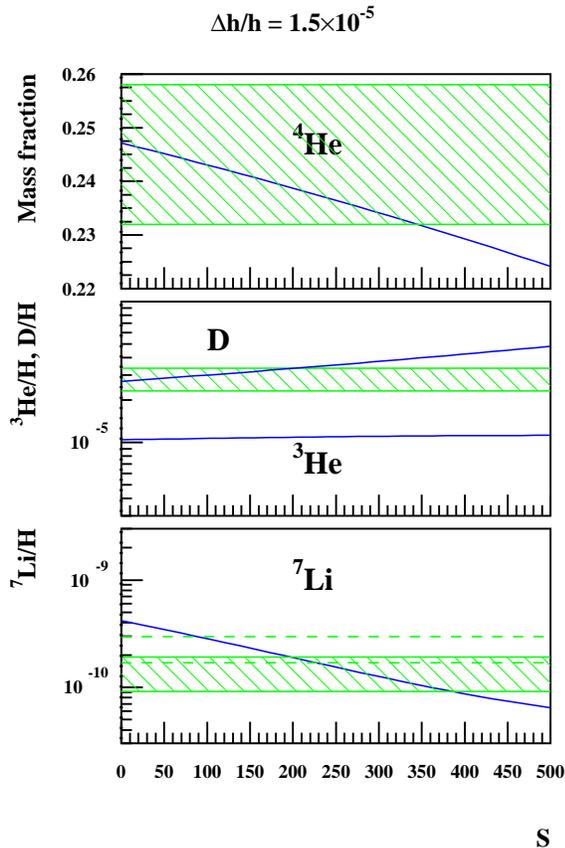}
\caption{Primordial abundances of the light nuclei as a function of the parameter $S$ assuming a change in the Yukawa couplings $\Delta h/h = 1.5 \times10^{-5}$ and $\Delta \alpha/\alpha = 0$.}
 \label{varS}
\end{figure}

Before concluding, we return once more to the question of model dependence. 
We have parametrized the uncertainty between $\Delta v$ and $\Delta h$ with the quantity $S$ and the uncertainty between $\Delta \Lambda$ and $\Delta \alpha$ through $R$. In full generality we ought to include one more unknown, say $T$, that parametrizes the relation between 
$\Delta \alpha$ and $\Delta h$, $T \equiv d\ln h/d\ln \alpha$ \cite{Langacker}. In this work, however, we focused our investigation in  the dilaton model where $T = 1/2$.
It is now important to evaluate more precisely how sensitive our results are to the value these parameters may take. In Fig.~\ref{varS} we illustrate the evolution of the primordial abundances of the light nuclei with $S$ for a fixed value of the change in the Yukawa couplings assuming $\Delta \alpha/\alpha = 0$. We clearly see that, in this case, the theoretical 
\li7 abundance  is compatible with its observational measurement provided 
$200 \lesssim S \lesssim 370$ (for the lower range of observational \li7 abundances).

\begin{figure}[htb]
\center
\includegraphics[width=8cm]{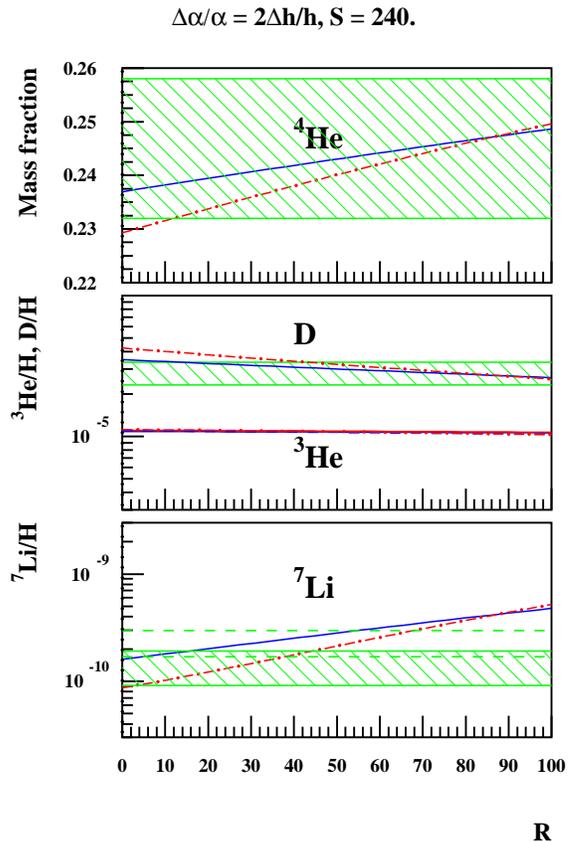}
\caption{Primordial abundances of the light nuclei as a function of the parameter $R$ assuming a change in the Yukawa couplings: $\Delta h/h = 1.5 \times10^{-5}$ (solid lines) and  $\Delta h/h = 2.5 \times10^{-5}$ (dot-dashed lines), for $S = 240$.}
 \label{varR}
\end{figure}

We can also evaluate the impact of changing $R$ in the dilaton model, when we allow a variation in $\alpha$. To this end we show in Fig.~\ref{varR} the evolution of the primordial abundances for two different values of $\Delta h/h$. 
We observe that when $\Delta h/h = 1.5 \times 10^{-5}$, we require $R = 16$. On the other hand, if we take $\Delta h/h = 2.5 \times 10^{-5}$,  the abundances are more sensitive to the value of $R$ as the
slope of the corresponding curves are steeper, but there is also a narrow window 
around $R = 45$ where all the light nuclei abundances are compatible with the full observational data.

\section{Summary}\label{sec5}

In this article, we have considered the influence of a possible variation of the fundamental
constants on the abundances of the light elements synthesized during BBN. We have focused our
attention on three fundamental quantities central to BBN, namely $Q$, $\tau_n$ and $B_D$, the
variation of which was related to the one of the fundamental constants. Specifying our theoretical
framework we have reduced the fundamental constants to two independent ones, the Yukawa coupling
and the fine structure constant.

We have shown that these constants have a strong effect on \he4 allowing us to set strong
constraints on the variation of $m_e$, $B_D$, $Q$ and $\tau_n$. Interestingly, the deuterium and
\li7 abundances are mainly sensitive to $B_D$ and we have shown that there is a window in
which \li7 is compatible with spectroscopic data (see Fig.~\ref{f:var_all}). The existence of
such a narrow window also implies that our mechanism can be falsified by an increase of the
precision of deuterium and/or \li7 data. Our analysis also enables one to set sharper constraints on the variation of the fundamental constants.

Assuming that the fine structure constant does not vary, we have shown that deuterium and \he4 data
set strong constraints on the variation of the Yukawa couplings [see Eq.~(\ref{limit})] but that
inside this bound there exists a range reconciling the \li7 abundance with spectroscopic
observations. We then allow the fine structure constant to vary and set a sharp constraint on its variation in the dilaton scenario [see Eqs.~(\ref{limit2a}) and (\ref{limit2b})]. The theoretical limitations have also been discussed in detail. More specifically, we have parametrized the relations between 
$\Delta v$ and $\Delta h$ and between $\Delta \Lambda$ and $\Delta \alpha$ with two free quantities, $S$ and $R$, respectively. We found that the specific value of these quantities plays an important role alongside the change in the fundamental parameters in solving the \li7 abundance problem. We conclude therefore, that a better understanding of the values of these parameters from the theoretical standpoint can help us to better constrain the variation of the fundamental parameters at the time of BBN.

\noindent{\bf Acknowledgements}: We would like to thank M. Voloshin for discussions.
The work of K.A.O. and N.J.N. was supported in part by DOE grant
DE--FG02--94ER--40823. The work is also supported by the project
``INSU--CNRS/USA''.

%%%%%%%%%%%%%%%%%%%%%%%%%%%%%%%%%%%%%%%%%%%%%%%%%%%%%%%%%%%%%%%%%%%%%%%%%%%%%%%%%%%%%%%%%


\begin{thebibliography}{99}

\bibitem{bbn}
 T. P. Walker, {\it et al.},
 Astrophys. J. {\bf 376}, 51 (1991);
 K. A. Olive, G. Steigman, and T. P. Walker,
 Phys. Rep. {\bf 333}, 389 (2000);
 B. D. Fields and S. Sarkar, in
 W.~M.~Yao {\it et al.}  [Particle Data Group],
  %``Review of particle physics,''
  J.\ Phys.\ G {\bf 33}, 220 (2006).
  %%CITATION = JPHGB,G33,1;%%.

\bibitem{bbnb}
 R.~H.~Cyburt, B.~D.~Fields and K.~A.~Olive,
 New Astron. {\bf 6} 215 (2001);
 A. Coc {\it et al.},
 Phys. Rev. D {\bf 65}, 043510 (2002).

\bibitem{wmap}
 D.~N.~Spergel {\it et al}.,
 [{\tt arXiv:astro-ph/0603449}].

\bibitem{cfo3}
 R.~H.~Cyburt, B.~D.~Fields and K.~A.~Olive,
 Phys.\ Lett.  B {\bf 567}, 227 (2003).

\bibitem{Coc04}
 A.~Coc {\it et al.},
 Astrophys.\ J.\ , {\bf 600}, 544 (2004).

\bibitem{cuoco}
 A.~Cuoco {\it et al.},
 Int.\ J.\ Mod.\ Phys.\ A {\bf 19}, 4431 (2004).

\bibitem{cyburt}
 R.~H.~Cyburt,
 Phys.\ Rev.\ D {\bf 70}, 023505 (2004).

\bibitem{rbofn}
 S.G. Ryan {\it et al.},
 Astrophys. J. Lett. {\bf 530}, L57 (2000).
 
 \bibitem{bona}
 P.~Bonifacio {\it et al.},
  %``First stars VII. Lithium in extremely metal poor dwarfs,''
  arXiv:astro-ph/0610245.
  %%CITATION = ASTRO-PH 0610245;%%

\bibitem{dep}
 S. Vauclair and C. Charbonnel,
 Astrophys. J. {\bf 502}, 372 (1998);
 M.~H.~Pinsonneault, {\it et al.},
 Astrophys. J.  {\bf 527}, 180 (1999);
 M.~H.~Pinsonneault, {\it et al.},
 Astrophys. J. {\bf 574}, 398 (2002);
 O.~Richard, G.~Michaud and J.~Richer,
 Astrophys. J. {\bf 619}, 538 (2005).

\bibitem{rnb}
 S.G. Ryan, J.E. Norris, and T.C. Beers,
 Astrophys. J. {\bf 523}, 654 (1999).

\bibitem{mr}
 J.~Melendez and I.~Ramirez,
 Astrophys. J.  {\bf 615}, L33 (2004).

\bibitem{Ang05}
 C. Angulo {\it et al.},
 Astrophys. J. {\bf 630}, L105 (2005).

\bibitem{cfo4}
 R.~H.~Cyburt, B.~D.~Fields and K.~A.~Olive,
 Phys. Rev. D {\bf 69}, 123519 (2004).

\bibitem{bah}
 J.N. Bahcall, M.H. Pinsonneault, and S. Basu,
 Astrophys. J. {\bf 555}, 990 (2001).

\bibitem{sno}
 S.~N.~Ahmed {\it et al.}  [SNO Collaboration],
 Phys. Rev. Lett. {\bf 92}, 181301 (2004).

\bibitem{cyburt2}
 R.~H.~Cyburt {\it et al.},
 Astropart. Phys. {\bf 23}, 313 (2005).

 \bibitem{inhomobbn}
 H.~Kurki-Suonio, {\it et al.},
 Astrophys.\ J.\  {\bf 353}, 406 (1990);
 K. Jedamzik and J.B. Rhem,
 Phys. Rev. D {\bf64}, 023510 (2001).

\bibitem{jed}
 K. Jedamzik,
 Phys. Rev. Lett. {\bf 84}, 3248 (2000);
 Phys. Rev. D {\bf 70}, 063524 (2004);
 Phys. Rev. D {\bf 70}, 083510 (2004).

\bibitem{susy}
 J.~L.~Feng, A.~Rajaraman and F.~Takayama,
 Phys. Rev. D {\bf 68}, 063504 (2003);
 J.~L.~Feng, S.~Su and F.~Takayama,
 Phys. Rev. D {\bf 70}, 063514 (2004);
 J.~L.~Feng, S.~Su and F.~Takayama,
 Phys.\ Rev.\ D {\bf 70}, 075019 (2004);
 J.~R.~Ellis, {\it et al.},
 Phys. Lett. B {\bf 588}, 7 (2004);
 J.~R.~Ellis, {\it et al.},
 Phys. Lett. B {\bf 573}, 162 (2003);
 J.~R.~Ellis, {\it et al.},
 Phys. Rev. D {\bf 70}, 055005 (2004);
 J.~R.~Ellis, K.~A.~Olive and E.~Vangioni,
 Phys. Lett. B {\bf 619}, 30 (2005);
 D.~G.~Cerdeno, {\it et al.},
 [{\tt arXiv:hep-ph/0509275}].

\bibitem{Jed2}
  K.~Jedamzik, {\it et al.},
  [{\tt arXiv:hep-ph/0512044}];
  R.~H.~Cyburt, {\it et al.},
  [{\tt arXiv:astro-ph/0608562}].

\bibitem{dn}
 T. Damour and K. Nordtvedt,
 Phys. Rev. Lett. {\bf70}, 2217 (1993);
 Phys. Rev. D {\bf48}, 3436 (1993).

\bibitem{dp}
 T. Damour and B. Pichon,
 Phys. Rev. D {\bf 59}, 123502 (1999);
 A.~Coc, {\it et al.},
 Phys. Rev. D {\bf 73}, 083525 (2006).

\bibitem{stother}
 C. Schimd, J.-P. Uzan, and A. Riazuelo
 Phys. Rev. D {\bf 71}, 083512 (2005);
 J. Martin, C. Schimd and J.-P. Uzan
 Phys. Rev. Lett. {\bf96}, 061303 (2006);
 A. Riazuelo and J.-P. Uzan,
 Phys. Rev. D {\bf 66}, 023525 (2002);
 J.-P. Uzan,
 Gen. Relativ. Gravit. {\bf 39}, 307 (2007).

\bibitem{bbna}
 E.~W.~Kolb, M.~J.~Perry and T.~P.~Walker,
 Phys. Rev. D {\bf33}, 869 (1986);
 L.~Bergstrom, S.~Iguri and H.~Rubinstein,
 Phys. Rev. D {\bf60}, 045005 (1999);
 K.~Ichikawa and M.~Kawasaki,
 Phys. Rev. D {\bf65}, 123511 (2002);
 K.M. Nollett and R.E. Lopez,
 Phys. Rev. D {\bf66}, 063507 (2002).

\bibitem{bbnb2}
  K.~Ichikawa and M.~Kawasaki,
  Phys. Rev. D {\bf 69}, 123506 (2004);
  C.~M.~Muller, G.~Schafer and C.~Wetterich,
  Phys. Rev. D {\bf 70}, 083504 (2004).

\bibitem{co}
 B.~A.~Campbell and K.~A.~Olive,
 Phys. Lett. {\bf B345}, 429 (1995).

\bibitem{flam1}
 V.~V.~Flambaum and E.~V.~Shuryak,
 Phys. Rev. D {\bf 65}, 103503 (2002).

\bibitem{fl1}
  V.~F.~Dmitriev, V.~V.~Flambaum and J.~K.~Webb,
  Phys. Rev. D {\bf 69}, 063506 (2004).

\bibitem{vvar}
  V.V. Dixit and M. Sher,
  Phys. Rev. D {\bf D37}, 1097 (1988);
  R.~J.~Scherrer and D.~N.~Spergel,
  Phys. Rev. D {\bf 47}, 4774 (1993);
  %J.~J.~Yoo and R.~J.~Scherrer,
  %Phys. Rev. D {\bf 67}, 043517 (2003);
  B.~Li and M.~C.~Chu,
  Phys. Rev. D {\bf 73}, 023509 (2006).

\bibitem{scherrer}
J.~J.~Yoo and R.~J.~Scherrer,
  Phys. Rev. D {\bf 67}, 043517 (2003);


\bibitem{lvar}
 J.~P.~Kneller and G.~C.~McLaughlin,
 Phys. Rev. D {\bf 68}, 103508 (2003).

\bibitem{otherbd}
 F.J. Dyson,
 Sci. Am. {\bf225}, 50 (1971);
 P.C.W. Davies,
 J. Phys. A {\bf 5}, 1296 (1972);
 J.D. Barrow,
 Phys. Rev. D {\bf35}, 1805 (1987);
 T. Pochet {\it et al.},
 Astron. Astrophys. {\bf 243}, 1 (1991).

\bibitem{webb}
 J.~K.~Webb, {\it et al.},
 Phys. Rev. Lett. {\bf 82}, 884 (1999);
 M.~T.~Murphy {\it et al.},
 Mon. Not. R. Astron. Soc. {\bf 327}, 1208 (2001);
 J.~K.~Webb {\it et al.},
 Phys. Rev. Lett. {\bf 87}, 091301 (2001);
 M.~T.~Murphy, {\it et al.},
 Mon. Not. R. Astron. Soc. {\bf 327}, 1223 (2001).

\bibitem{murphy3}
 M.~T.~Murphy, J.~K.~Webb and V.~V.~Flambaum,
 [{\tt arXiv:astro-ph/0306483}].

\bibitem{chand}
 H.~Chand, {\it et al.},
 Astron. Astrophys.  {\bf 417}, 853 (2004);
 R.~Srianand, {\it et al.},
 Phys. Rev. Lett.  {\bf 92}, 121302 (2004).

\bibitem{amo}
 T.~Ashenfelter, G.~J.~Mathews and K.~A.~Olive,
 Phys. Rev. Lett. {\bf 92}, 041102 (2004);
 T.~P.~Ashenfelter, G.~J.~Mathews and K.~A.~Olive,
 Astrophys. J.  {\bf 615}, 82 (2004);
 M.~G.~Kozlov, {\it et al.},
 Phys. Rev. A {\bf 70}, 062108 (2004).

\bibitem{mu0}
 A.~Y.~Potekhin, {\it et al.},
 Astrophys. J.  {\bf 505}, 523 (1998);
 A.~V.~Ivanchik, {\it et al.},
 Astron. Lett. {\bf 28}, 423 (2002);
 S.~A.~Levshakov, {\it et al.},
 Mon. Not. R. Astron. Soc. {\bf 333}, 373 (2002).

\bibitem{mu}
 A.~Ivanchik {\it et al.},
 Astron. Astrophys.  {\bf 440}, 45 (2005);
 E.~Reinhold, {\it et al.},
 Phys. Rev. Lett. {\bf 96}, 151101 (2006).

\bibitem{ctes}
 J.-P. Uzan,
 Rev. Mod. Phys. {\bf 75}, 403 (2003);
 AIP Conf. Proceedings {\bf 736} (2004) 3,
 [{\tt arXiv:astro-ph/0409424}];
 G.F.R. Ellis and J.-P. Uzan,
 Am. J. Phys. {\bf73}, 240 (2005).

\bibitem{landau}
 S.~J.~Landau, M.~E.~Mosquera and H.~Vucetich,
 Astrophys. J. {\bf 637}, 38 (2006).

\bibitem{bbng}
 F.S. Accetta, L.M. Krauss, and P. Romanelli,
 Phys. Lett. B {\bf248}, 146 (1990);
 C.~J.~Copi, A.~N.~Davis and L.~M.~Krauss,
 Phys. Rev. Lett. {\bf 92}, 171301 (2004);
 J.A. Casas, J. Garcia-Bellido, and N. Quiros,
 Mod. Phys. Lett. A {\bf7}, 447 (1992);
 T. Damour and C. Gundlach,
 Phys. Rev. D {\bf43}, 3873 (1991).

\bibitem{weaktran}
 K. Inoue, {\it et al.},
 Prog. Theor. Phys. {\bf 68}, 927 (1982);
 L.E. Ib\'{a}\~{n}ez and G.G. Ross,
 Phys. Lett. {\bf B110}, 215 (1982);
 J. Ellis, L.E. Ib\'{a}\~{n}ez and G.G. Ross,
 Phys. Lett. {\bf B113}, 283 (1982);
 L. Alvarez-Gaum\'{e}, M. Claudson, and M. Wise,
 Nucl. Phys. {\bf B207}, 96 (1982) 96;
 J. Ellis, {\it et al.},
 Phys. Lett. {\bf B125}, 275 (1983).
 
 \bibitem{opqccv}
 K.~A.~Olive, {\it et al.},
  Phys.\ Rev.\ D {\bf 66}, 045022 (2002);
  K.~A.~Olive, {\it et al.},  
  Phys.\ Rev.\  D {\bf 69}, 027701 (2004).

\bibitem{pn}
 J. Gasser and H. Leutwyler,
 Phys. Rep. {\bf 87}, 77 (1982).

\bibitem{Audi}
 G.~Audi, A.H.~Wapstra and C.~Thibault,
 Nucl. Phys. {\bf A729}, 337 (2003).

 \bibitem{fl}
 V.~F.~Dmitriev and V.~V.~Flambaum,
 Phys. Rev. D {\bf 67}, 063513 (2003);
 V.~V.~Flambaum and E.~V.~Shuryak,
 Phys. Rev. D {\bf 67}, 083507 (2003).

\bibitem{oldsnp}
 J. Gasser, H. Leutwyler, and M. E. Sainio,
 Phys. Lett. {\bf B253}, 252 (1991);
 M.~Knecht,
 [{\tt arXiv:hep-ph/9912443}];
 M.~E.~Sainio,
 PiN Newslett. {\bf 16}, 138 (2002).

\bibitem{Cheng}
 H.-Y. Cheng,
 Phys. Lett. {\bf B219}, 347 (1989).

\bibitem{eoss8}
  J.~R.~Ellis, {\it et al.},
  Phys. Rev. D {\bf 71}, 095007 (2005).

\bibitem{highsnp}
 M.~M.~Pavan, {\it et al.},
 PiN Newslett.\  {\bf 16}, 110 (2002).

\bibitem{exotic}
 P.~Schweitzer,
 Eur.\ Phys.\ J.\ A {\bf 22}, 89 (2004);
 J.~R.~Ellis, M.~Karliner and M.~Praszalowicz,
 JHEP {\bf 0405}, 002 (2004).

\bibitem{svz}
 M.\ A.\ Shifman, A.\ I.\ Vainshtein and V.\ I.\ Zakharov,
 Phys.\ Lett. B {\bf 78}, 443 (1978);
 A.\ I.\ Vainshtein, V.\ I.\ Zakharov and M.\ A.\ Shifman,
 Usp.\ Fiz.\ Nauk {\bf 131}, 537 (1980).

\bibitem{Dent}
  T.~Dent and M.~Fairbairn,
  Nucl.\ Phys.\ B {\bf 653}, 256 (2003)
  T.~Dent,
  Nucl.\ Phys.\ B {\bf 677}, 471 (2004).

\bibitem{dine}
 M.~Dine, Y.~Nir, G.~Raz and T.~Volansky,
 Phys.\ Rev.\ D {\bf 67}, 015009 (2003).

\bibitem{Langacker}
 P.~Langacker, G.~Segre and M.~J.~Strassler,
 Phys.\ Lett.\ B {\bf 528}, 121 (2002).

\bibitem{cmssm}
 J.~R.~Ellis, G.~Ganis and K.~A.~Olive,
 Phys.\ Lett.\ B {\bf 474}, 314 (2000);
 V.~D.~Barger and C.~Kao,
 Phys.\ Lett.\ B {\bf 518}, 117 (2001);
 L.~Roszkowski, R.~Ruiz de Austri and T.~Nihei,
 JHEP {\bf 0108}, 024 (2001);
 A.~Djouadi, M.~Drees and J.~L.~Kneur,
 JHEP {\bf 0108}, 055 (2001);
 J.~R.~Ellis, {\it et al.},
 Phys. Lett. B {\bf565}, 176 (2003);
 H.~Baer and C.~Balazs,
 JCAP {\bf 0305}, 006 (2003);
 A.~B.~Lahanas and D.~V.~Nanopoulos,
 Phys. Lett. B {\bf 568}, 55 (2003);
 U.~Chattopadhyay, A.~Corsetti and P.~Nath,
 Phys. Rev. D {\bf68}, 035005 (2003);
 C.~Munoz,
 Int.\ J.\ Mod.\ Phys.\ {\bf A19}, 3093 (2004);
 R.~Arnowitt, B.~Dutta and B.~Hu,
 [{\tt arXiv:hep-ph/0310103}].


\bibitem{EENZ}
 J.~Ellis, {\it et al.},
 Mod.\ Phys.\ Lett.\ A {\bf 1}, 57 (1986);
 R.~Barbieri and G.~F.~Giudice,
 Nucl.\ Phys.\ B {\bf 306}, 63 (1988).

\bibitem{eos}
 J.~R.~Ellis, K.~A.~Olive and Y.~Santoso,
 New Jour.\ Phys.\  {\bf 4}, 32 (2002).

\bibitem{epelbaum}
  E.~Epelbaum, U.~G.~Meissner and W.~Gloeckle,
  %``Nuclear forces in the chiral limit,''
  Nucl.\ Phys.\  A {\bf 714}, 535 (2003);
  S.~R.~Beane and M.~J.~Savage,
  %``The quark mass dependence of two-nucleon systems,''
  Nucl.\ Phys.\  A {\bf 717}, 91 (2003)

\bibitem{Des04}
 P.~Descouvemont,{\it et al.},
 ADNDT, {\bf 88}, 203 (2004).

\bibitem{Che99}
 J.-W. Chen and M. Savage,
 Phys. Rev. C {\bf60}, 065205 (1999).

\bibitem{Dic82}
 D.~Dicus, {\it et al.},
 Phys. Rev. {\bf D 26}, 2694 (1982).

\bibitem{Lop99}
 R. Lopez and M. Turner,
 Phys. Rev. {\bf D 59}, 103502 (1999).

\bibitem{kneller2004}
  J.~P.~Kneller and G.~C.~McLaughlin,
  %``The Effect of Bound Dineutrons upon BBN,''
  Phys.\ Rev.\  D {\bf 70}, 043512 (2004).

\bibitem{Rupak}
  G.~Rupak, T.~Schafer and A.~Kryjevski,
  %``Polarized fermions in the unitarity limit,''
  arXiv:cond-mat/0607834.

\bibitem{PDG06}
 W.~M.~Yao {\it et al.}  [Particle Data Group],
 J.\ Phys.\ G {\bf 33}, 1 (2006).
 
 \bibitem{newn}
 A.~Serebrov {\it et al.},
  %``Measurement of the neutron lifetime using a gravitational trap and a
  %low-temperature Fomblin coating,''
  Phys.\ Lett.\  B {\bf 605}, 72 (2005)
  [arXiv:nucl-ex/0408009].
  %%CITATION = PHLTA,B605,72;%%

\bibitem{mks}
G.~J.~Mathews, T.~Kajino and T.~Shima,
  %``Big Bang Nucleosynthesis with a New Neutron Lifetime,''
  Phys.\ Rev.\  D {\bf 71}, 021302 (2005)
  [arXiv:astro-ph/0408523].
  %%CITATION = PHRVA,D71,021302;%%

\bibitem{os}
  K.~A.~Olive and E.~D.~Skillman,
  Astrophys.\ J.\  {\bf 617}, 29 (2004).

\bibitem{omear}
  J.~M.~O'Meara, {\it et al.},
  [{\tt arXiv:astro-ph/0608302}].

\end{thebibliography}
\end{document}